\documentclass[aip,rsi,reprint,graphicx]{revtex4-1} 

\usepackage{graphicx}

\draft 

\begin{document}


\title{PHARAO Laser Source Flight Model: Design and Performances} 



\author{T. L\'ev\`eque}
\email[]{Electronic mail: thomas.leveque@cnes.fr}
\author{B. Faure}
\author{F.X. Esnault}
\author{C. Delaroche}
\author{D. Massonnet}
\author{O. Grosjean}
\author{F. Buffe}
\author{P. Torresi}

\affiliation{Centre National d'Etudes Spatiales, 18 avenue Edouard Belin, 31400 Toulouse, France.}

\author{T. Bomer}
\author{A. Pichon}
\author{P. B\'eraud}
\author{J. P. Lelay}
\author{S. Thomin}
\affiliation{Sodern, 20 Avenue Descartes, 94451 Limeil-Br\'evannes, France.}

\author{Ph. Laurent}
\affiliation{LNE-SYRTE, Observatoire de Paris, CNRS, UPMC, 61 avenue de l'Observatoire, 75014 Paris, France.}

\date{\today}

\begin{abstract}
In this paper, we describe the design and the main performances of the PHARAO laser source flight model. PHARAO is a
laser cooled cesium clock specially designed for operation in space and the laser source is one of the main sub-systems. The flight model presented in this work is the first remote-controlled laser system designed for spaceborne cold atom manipulation. The main challenges arise from mechanical compatibility with space constraints, which impose a high level of compactness, a low electric power consumption, a wide range of operating temperature and a vacuum environment. We describe the main functions of the laser source and give an overview of the main technologies developed for this instrument. We present some results of the qualification process. The characteristics of the laser source flight model, and their impact on the clock performances, have been verified in operational conditions.
\end{abstract}


\maketitle 

\section{Introduction}

PHARAO (Projet d'Horloge Atomique par Refroidissement d'Atomes en Orbite), is the first cold atom clock specially designed for operation in space~\cite{laurent2006}. This development is managed by the French space agency, CNES, within the framework of the ACES (Atomic Clock Ensemble in Space) mission. ACES is a European space mission managed by the European Space Agency (ESA). The mission is dedicated to fundamental physics through the time and frequency comparisons between the space clock and ground based clocks~\cite{salomon2001, cacciapuoti2009}. The clock signal of the space segment is generated from the combination of the PHARAO clock and an active hydrogen maser~\cite{goujon2010, moric2013}. The expected performances in space are frequency accuracy better than 3$\times 10^{-16}$ and frequency stability of 10$^{-13} \tau^{-1/2}$, where $\tau$ is the integration time in seconds. The ACES payload will be installed on the external support of the European Columbus laboratory aboard the International Space Station (ISS). Ground clocks will be located in Europe, USA, Japan and Australia and the mission will be operated from the ACES user support and operation center, located at CNES in France. The launch is scheduled for 2017. Planned duration of the mission is 18-36 months.

\begin{figure}[!h]
 \includegraphics[width=8.5cm]{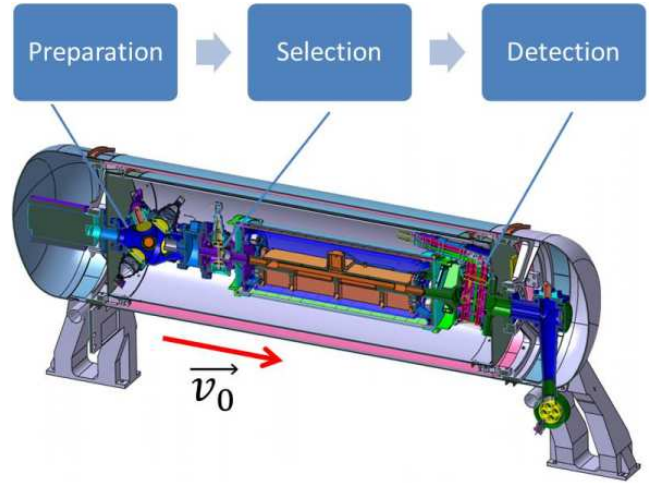}
 \caption{\label{tc_pharao}Sectional view of the PHARAO cesium tube where the cold atoms are manipulated. The interaction areas between the cesium atoms and the lasers for the preparation, selection and detection processes are highlighted. The direction of the launching velocity of the cold atoms $\mathbf{v_0}$ is indicated by the red arrow.}
 \end{figure}

In the PHARAO clock the cesium atoms are manipulated in a high vacuum chamber called cesium tube (Fig.~\ref{tc_pharao}). The laser source provides the laser beams to the cesium tube in order to prepare, select and detect the atoms. The microwave source provides two 9.2~GHz signals to the cesium tube and the corrected metrological 100~MHz signal to the ACES payload. A computer drives the clock operation, receives telecommands from the ground segment and transfers operational telemetry. These four sub-systems have been space qualified and the flight model of the PHARAO clock is now assembled and tested.

In this paper, we focus on the laser source design, technologies and performances. The laser source is a complex system made of free space optics which must be aligned to handle the laser beams while keeping a high level of stability. The main challenges also arise from the mechanical compatibility with space launch: high level of compactness, low electric power consumption, a wide range of environment temperatures and under vacuum operation. The functions and the main requirements of the laser source are presented in Sec.~\ref{sec_requirements}. The Sec.~\ref{sec_design} describes the laser set-up and shows the key technologies. Finally, the experimental results are presented in the Sec.~\ref{sec_perfo}.

\section{Laser source requirements}
\label{sec_requirements}

\subsection{General requirements}

The laser source has three main functions during  the PHARAO clock operation: the cold atom preparation, the atomic state selection and the atomic state detection. The different interaction areas inside the cesium tube between the atoms and the lasers beams  are depicted in the Fig.~\ref{tc_pharao}. Two main laser frequencies are needed to manipulate the cesium atoms. The first one, called "Master frequency", is tuned around the $|6S_{1/2},F = 4 \rangle$ $\rightarrow$ $|6P_{3/2},F' = 5 \rangle$ cesium cycling transition~\cite{steck2010}. The second one, called "Pump frequency", is locked on the $|6S_{1/2},F = 3 \rangle$ $\rightarrow$ $|6P_{3/2},F' = 4 \rangle$ transition. The cold atoms are manipulated by six laser beams made of two independently controllable triplets called "Lower beams" and "Upper beams".

\subsection{Cold atom preparation}

\begin{figure}[!h]
 \includegraphics*[width=8.5cm]{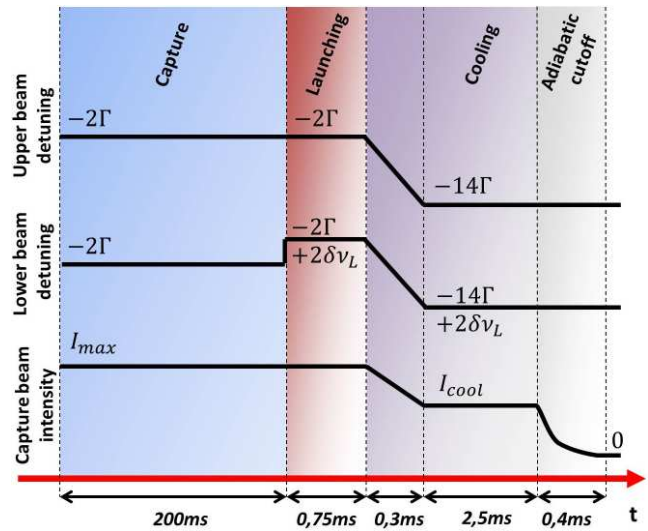}
 \caption{\label{preparation} Typical sequence of atom preparation. $2\delta \nu_L$ is the lower cooling beam detuning which defines the launching velocity. $\Gamma$=5.3 MHz is the natural linewidth of the $^{133}$Cs D2 line. $I_{\mathrm{max}}$ is the laser beam intensity used during the capture and launching phase. $I_{\mathrm{cool}}$ is the laser beam intensity used for the cooling phase.}
\end{figure}

The cold atom preparation requires the power and frequencies of the lasers to be changed in a specific sequence as described in Fig.~\ref{preparation}. First the laser beams capture and cool cesium atoms coming from a thermal vapor to form the so-called optical molasses~\cite{dalibard1989}. The vapor pressure of cesium at the level of the preparation area is of 2$\times 10^{-6}$~Pa. The number of captured atoms is a function of the master laser power and detuning. Thus, in order to fulfill the mission requirements, we consider that a power of 12~mW per beam must be available while the frequency can be red-detuned up to 10~MHz below the cycling transition. The capture time can be changed from 200~ms to several seconds in order to  optimize the clock operation. During the cold atom preparation 5~mW of pump laser power are required to cool the atoms efficiently.  

The cold atoms are then launched by using the moving molasses technique. One triplet is frequency shifted by 2$\delta \nu_L$ to reach a velocity $v_0$ given by $\sqrt{3} \delta \nu_L\lambda$. The launching velocity can be chosen between 0.05 and 5~m/s with a step of 0.05~m/s. The launch duration can be adjusted from 100~$\mu$s to few ms depending on the launching velocity. In order to cool down the atoms, the laser power is then ramped down and the laser detuning is increased. Finally, the laser power is switched off by following an adjustable exponential decay. The powers of the 6 laser beams must be balanced within few percents in order to guarantee the isotropy of the preparation process. These powers are measured inside the cesium tube.

\subsection{Cold atom selection}

Following the preparation, the atoms are distributed among the nine Zeeman sublevels of the $|6S_{1/2},F = 4 \rangle$ state. The operation of the clock requires the atoms to be only in the magnetic field  insensitive state $|6S_{1/2},F = 3, m_{F}=0 \rangle$. For this purpose, a static magnetic field is applied to lift the degeneracy of the Zeeman sublevels. Atoms in $|6S_{1/2},F = 4, m_{F}=0 \rangle$ are transferred to $|6S_{1/2},F = 3, m_{F}=0 \rangle$ when passing through a microwave cavity. 

The selection uses two laser beams. The first one pumps part of the cloud from F=3 to F= 4. The second one removes all the atoms remaining in F=4 by radiation pressure. This combination allows to slice the cloud (particularly useful at low launch velocity) and to keep atoms in a pure quantum state.

Two optical fibers, carrying respectively the Master frequency and the Pump frequency lasers provide the beams for the cold atom selection. The power, the frequency, and the illumination duration of each beam are adjustable. The maximum powers at the output of the master and pump fibers are respectively of 1.5~mW and 400~$\mu$W.

\begin{figure*}[htb]
 \includegraphics[width=17cm]{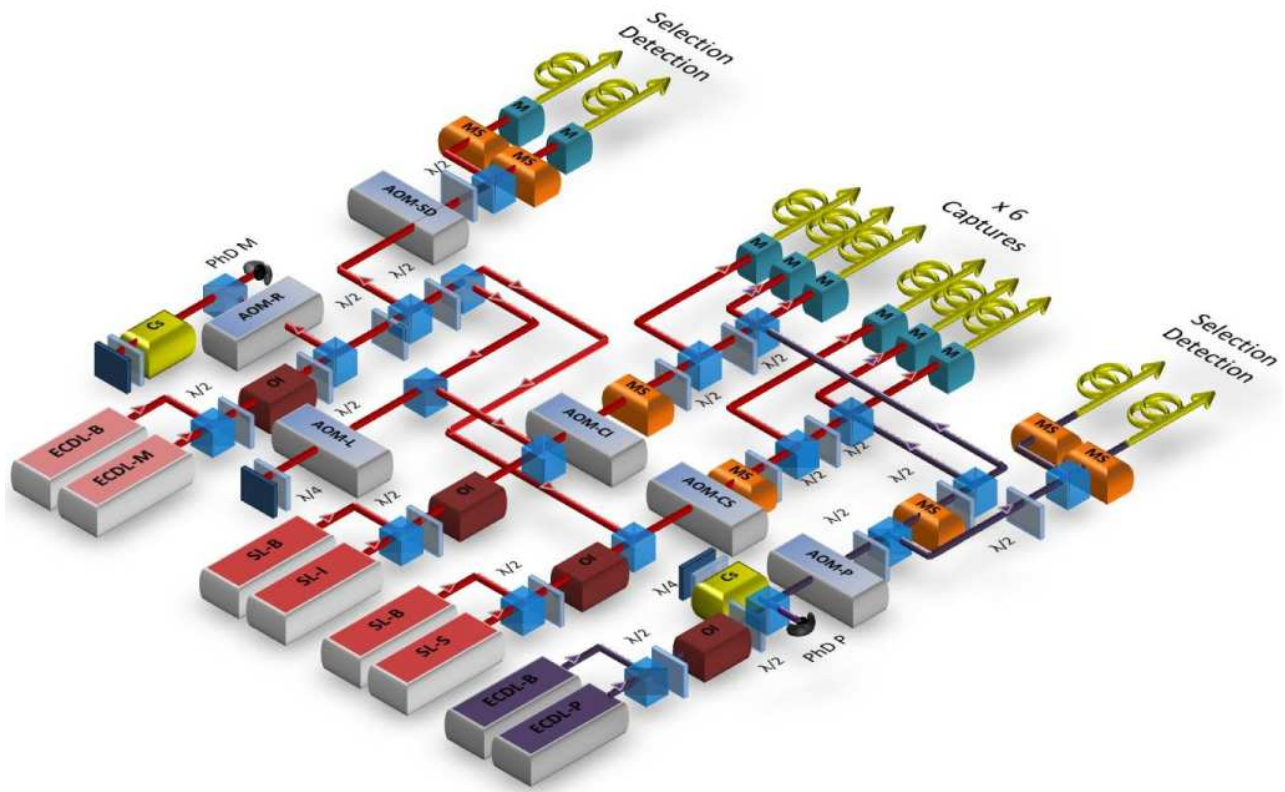}
 \caption{\label{SL_setup} Diagram of the complete laser system. ECDL extended cavity laser diodes at 852~nm, SL slave lasers, OI optical isolators,  AOM acousto-optic modulator, MS mechanical shutters, M piezoelectric mirrors, Cs Cesium cells. The optical fibers are represented by the yellow arrows. The mechanisms used for switching from the nominal to the backup lasers (ECDL-B, SL-B) are not represented.}
 \end{figure*}

\subsection{Cold atom detection}

After selection in the $|6S_{1/2},F = 3, m_{F}=0 \rangle$ state, atoms undergo a Ramsey interaction inside a microwave cavity which transfers them in the  $|6S_{1/2},F = 4, m_{F}=0 \rangle$ state with a probability $P$. This transition probability is measured to derive the error signal and frequency corrections to be applied to the microwave source. The detection process consists in measuring the probability $P$, defined by the ratio N4/(N3+N4) where N3 (resp. N4) is the number of atoms in the $|6S_{1/2},F = 3\rangle$ (resp. $|6S_{1/2},F = 4\rangle$) state.  After the Ramsey interrogation, the atoms pass through a first master probe beam. The fluorescence light emitted by the cycling transition is collected on a photodiode. These detected atoms are then eliminated by a master pusher beam similar to the one used in the selection process. The remaining cloud, containing only atoms in the $F$=3 state, is optically pumped to the $F$=4 state while passing through a pumping beam. These atoms are finally detected by fluorescence with another master beam.

The Master and the Pump lasers are provided to the cesium tube through two optical fibers. In order to perform the detection process, the maximum powers available at the output of the master and pump fibers are respectively of 10~mW and 94~$\mu$W.

The relative frequency and intensity fluctuations of the lasers can induce variations on the spontaneous emission rate during the detection process and add noise to the clock error signal. To reduce this noise source, the laser frequency and the laser intensity are locked on a reference. The specification of the laser noise, depicted on the Fig.~\ref{DSP}, has been determined in order to make this contribution negligible on the clock frequency stability.

\section{Laser source design}
\label{sec_design}

\subsection{Laser setup}

The laser source architecture~\cite{laurent2006} is shown in Fig.~\ref{SL_setup}. The frequency of the Master extended cavity diode laser (ECDL-M) at 852~nm is stabilized to the saturated absorption crossover resonance of cesium ($|6S_{1/2},F = 4 \rangle$ $\rightarrow$ $|6P_{3/2},F' = 4/5 \rangle$) through an acousto-optic modulator (AOM-R). To perform the frequency lock, the diode laser current is modulated at 500 kHz with a modulation index of 4. The absorption signal is demodulated to provide the error signal which is integrated and applied to the diode laser current and to the PZT transducer that controls the cavity length. By changing the in-loop AOM-R rf driving frequency the laser frequency can be tuned over 60~MHz with a rate of about 200~MHz/ms. This ECDL/AOM combination provides the frequency reference for all the atom manipulation processes: capture, cooling, selection, and detection.

The ECDL-M beam is optically isolated by 37~dB and divided into three beams. One beam is coupled into two optical fibers through a 75~MHz acousto-optic modulator (AOM-SD) to adjust the power for the atomic selection (1.5mW) and detection (10mW). The detection laser power is measured inside the cesium tube and stabilized by acting on the AOM-SD rf driving power within a 20~kHz bandwidth.

The second beam passes through an acousto-optic modulator (AOM-L) twice before injecting a slave diode laser SL-I. The AOM-L frequency can be adjusted from 87 to 90~MHz to launch atoms (from 5~cm/s to 5~m/s). The third beam injects a second slave diode laser SL-S. The slave diode laser output power is of 100 mW for an input current of 140~mA. After optical isolation, slave laser beams are shifted by 90~MHz by two acousto-optic modulators (AOM-CI and AOM-CS) and split into six beams. The three beams coming from the SL-I slave laser are used to generate the lower capture beams while the three beams coming from SL-S are used for the upper ones. The beams are injected into six optical fibers by using a mirror mounted on PZT transducers (see Sec.~\ref{sec_MEF}). The PZT mirrors allow a fine alignment of the beam on the fiber core and a power balancing in each of the three molasses beam pairs. The laser powers measured at the output of the fibers, inside the cesium tube, are used to feed a servo-loop acting on the AOM-CI and AOM-CS in order to stabilize the capture beam intensities during the different preparation steps.

A second ECDL-P is locked to the cesium crossover resonance $|6S_{1/2},F = 3 \rangle$ $\rightarrow$ $|6P_{3/2},F' = 3/4 \rangle$. The beam passes through a 101~MHz acousto-optic modulator (AOM-P) and is split into four. Two beams are injected into optical fibers to repump the atoms during selection and detection. The two others are superimposed on four capture beams to repump the atoms during the preparation process.

\begin{figure}[!h]
 \includegraphics[width=8.5cm]{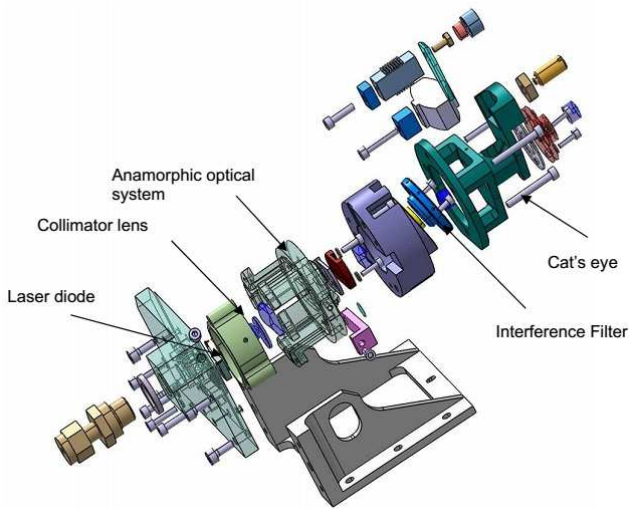}
 \caption{\label{LCE1}Exploded view of the ECDL mechanical design.}
 \end{figure}

The AOMs allow to switch on/off the beams by 60~dB with a risetime of about 20~$\mu$s. In addition, seven mechanical shutters (MS) ensure a complete extinction of the beams ($>$120~dB). ECDLs and Slave lasers have redundant models (ECDL-B, SL-B). The beams of the redundant lasers are superimposed onto the firsts by means of polarizing cubes. For switching from the nominal to the redundant laser, a $\lambda/2$ waveplate is inserted inside the beam before passing through the optical isolator. The setup is composed of 4 nominal lasers (ECDL-M, ECDL-P, SL-I, SL-S) and each laser has a redundant model. The system then offers 2$^4$=16 working configurations. The main difficulty arises from the optical alignment process and required mechanical stability to be able to switch among the 16 available configurations while keeping the same laser properties at the end.  

\begin{figure}[!h]
 \includegraphics[width=8.5cm]{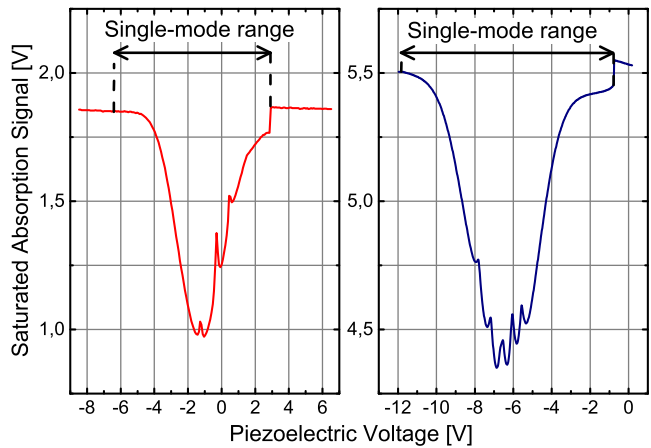}
 \caption{\label{abs_sat}\textit{Left:} Saturated absorption spectroscopy signal obtained with the ECDL-M tuned on the $|6S_{1/2},F = 4 \rangle$ line of the Cesium. \textit{Right:} Saturated absorption spectroscopy signal obtained with the ECDL-P tuned on the  $|6S_{1/2},F = 3 \rangle$ line of the Cesium.}
 \end{figure}

\subsection{Laser source technologies}

\subsubsection{Extended cavity laser diodes (ECDL)}
\label{sec_LCE}

The ECDL consists in an extended cavity laser diode in a linear configuration where the mode selection is performed by an intra-cavity interference filter~\cite{baillard2006,simon2004}. This design enables a rugged and reliable behavior and provides a spectral purity 10 to 100 times higher than simple laser diodes. The complete mechanical assembly is depicted in Fig.~\ref{LCE1}. The laser diode is a JDS 5421 in a TO 56 package hermetically sealed with dry air to prevent failure by laser induced contamination~\cite{riede2011}. The laser diode is thermally regulated by a Peltier element. At the output of the laser diode, the beam is collimated and shaped by an anamorphic optical system made of cylindrical lenses. The mode selection relies on an interference filter which angle is set to define the wavelength~\cite{baillard2006}. The external cavity is closed by a cat's eye configuration to ensure the stability of the alignment. The cavity length is of 6.5~cm. The mirror of the cat's eye has a reflectivity of 25\% and is mounted on a PZT translator to change the cavity length and tune the laser frequency.

The ECDL achieves a linewidth of 100~kHz and a single-mode tunability of 1~GHz. The Fig.~\ref{abs_sat} shows the saturated absorption spectroscopy of the D2 line of the Cs realized with the ECDL-M and ECDL-P lasers. The frequency locks are achieved with a feedback to the diode current (30~kHz bandwidth) and the length of the external cavity (100~Hz bandwidth). During operation, all the laser parameters and the feedback loop are controlled by the computer. It permanently checks the servoloop and relocks the lasers if necessary~\cite{allard2004}.

The nominal output power of the laser is of 30~mW for a current of 100~mA. The laser threshold is reached for a current of 13~mA. The power of the laser spurious side modes have been verified to be below -87~dBc up to 10~GHz from the carrier in a 3~MHz bandwidth.

 \begin{figure}[!h]
 \includegraphics[width=8.5cm]{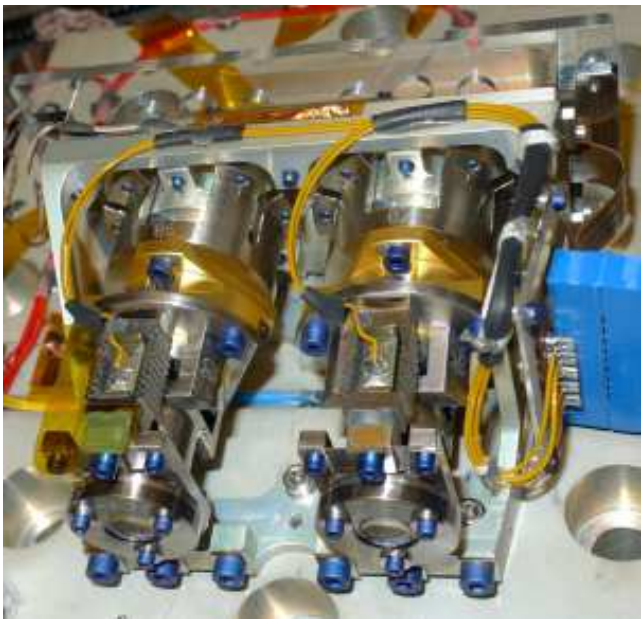}
 \caption{\label{LCE2}Photograph of the ECDL flight model module. This module includes two identical ECDL: the nominal laser (left) and its redundancy (right). The dimensions of the complete module are 80~mm x 100~mm x 45~mm.}
 \end{figure}

A picture of the ECDL module is given in Fig.~\ref{LCE2}. This module includes two identical ECDL: the nominal laser and its redundancy. The laser modules, and the mechanism for switching from the nominal to the redundant laser, have been qualified for space environment. In non-operational conditions, they have been submitted to thermal variations (-40$^{\circ}$C / +60$^{\circ}$C ) and random vibrations up to 40~$g_{\mathrm{rms}}$. No performances evolution have been observed after these tests.

\begin{figure}[!h]
 \includegraphics[width=8.5cm]{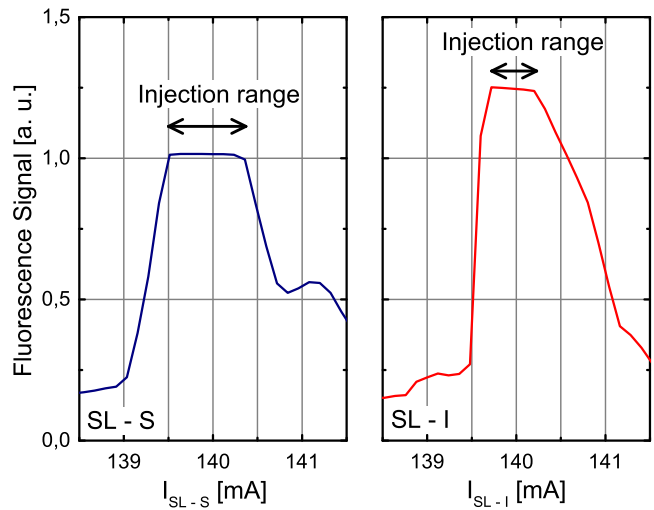}
 \caption{\label{scan_le}Fluorescence signal recorded when scanning the SL-S (left) and SL-I (right) currents. The injection ranges of the slave lasers are highlighted.}
 \end{figure}

In normal operation, the ECDL-M is locked and seeds the two slave laser diodes. Only 50~$\mu$W are required to inject the slave lasers. The quality of the injection of the slave laser is checked by monitoring the fluorescence signal of a cesium cell. The high fluorescence level indicates that the slave laser frequency is properly injected by the master laser. The Fig.~\ref{scan_le} represents the fluorescence signal as a function of the current setpoint of the slave lasers. The slave diode lasers remain injected over a 0.5~mA current variation. In regular operation, the temperature of the slave lasers is automatically set in order to center the fluorescence range on the current setpoint (140~mA). This ensures the stability of the injection process.

\subsubsection{Acousto-optic modulators}

\begin{figure}[!h]
 \includegraphics[width=8.5cm]{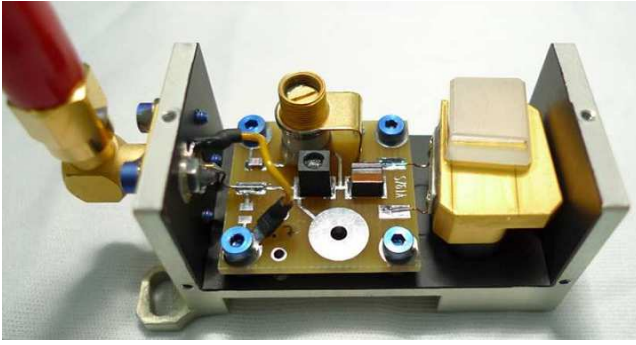}
 \caption{\label{MAO}Photograph of the acousto-optic modulator design.}
 \end{figure}

The acousto-optic modulators have two main functions in the laser setup. First, they shift and tune the frequency of the laser beams. Second, they control the laser power sent on each set of fibers. The device, shown in Fig.~\ref{MAO}, has been specially designed by Sodern and A\&A opto-electronic to be fully compatible with space applications. In particular, it has been designed to minimize power consumption. Thus the flight-model components achieve a diffraction efficiency of 90\% for a driving radio-frequency power of 200~mW. Moreover, the crystal shape has been designed to reduce the angle between the diffracted and incident beams to less than 1$^{\circ}$, simplifying the design and integration of the laser setup.

The acousto-optic modulator has been qualified for space environment. In non-operational conditions, the tests have proven a compatibility with thermal variations (-40$^{\circ}$C / +40$^{\circ}$C ) and random vibrations up to 40~$g_{\mathrm{rms}}$. Nevertheless, a risk of degradation of the AOM transducer has been observed for temperatures exceeding 40$^{\circ}$C.

\subsubsection{Mechanical shutters}

\begin{figure}[!h]
 \includegraphics[width=8.5cm]{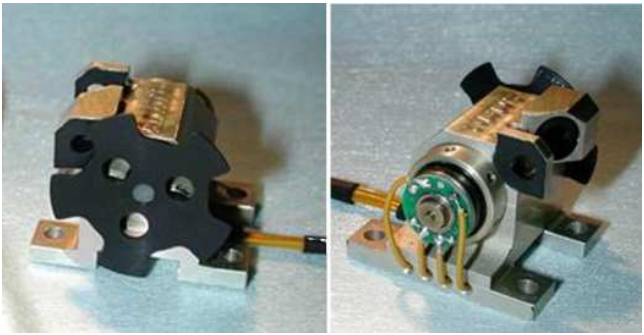}
 \caption{\label{shutter}Photograph of the mechanical shutter based on a paddle wheel mounted on a stepper motor.}
 \end{figure}

Mechanical shutters ensure an extinction better than 120~dB in 5~ms on the laser power. This function is essential to prevent any stray light inside the cesium tube during the Ramsey interrogation. The shutter design, depicted in Fig.~\ref{shutter} is based on the use of a paddle wheel mounted on a stepper motor. The same motors are also used for the redundancy mechanism described in Sec.~\ref{sec_LCE}.

The extinction ratio measurement was achieved by a fibered photon counter (ID Quantique, Ref. ID100-MMF50-ULN) calibrated on a low-level radiometer. With our system, the required extinction (120~dB) corresponds to a count rate of 200~Hz. The measurement, which includes the dark count rate, is less than 35~Hz. The device has been qualified to operate up to $2.18 \times 10^8$ cycles in space environment, in agreement with the mission lifetime. In non-operational conditions, the tests have proven a compatibility with thermal variations (-50$^{\circ}$C / +75$^{\circ}$C ) and random vibrations up to 30~$g_{\mathrm{rms}}$. This element has no magnetic effects on the clock as the cesium tube is surrounded by 3 layers of mu-metal magnetic shields.

\subsubsection{Piezoelectric mirrors}
\label{sec_MEF}

At the output of the laser setup, the beams are injected in 10 single mode polarization maintaining fibers. For 8 of these fibers, the injection is controlled by a two axis rotating mirror. These tip-tilt mechanisms are used to balance the powers of the 6 different capture laser beams at the percent level. Moreover, they allow the injection of the laser beams into the optical fibers to be periodically re-optimized during the mission if necessary. 

 \begin{figure}[!h]
 \includegraphics[width=8.5cm]{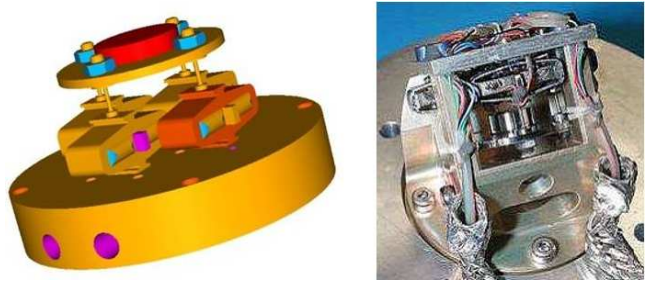}
 \caption{\label{MEF}Views of the piezoelectric mirror tip-tilt mechanism.}
 \end{figure}

 \begin{figure}[!h]
 \includegraphics[width=8.5cm]{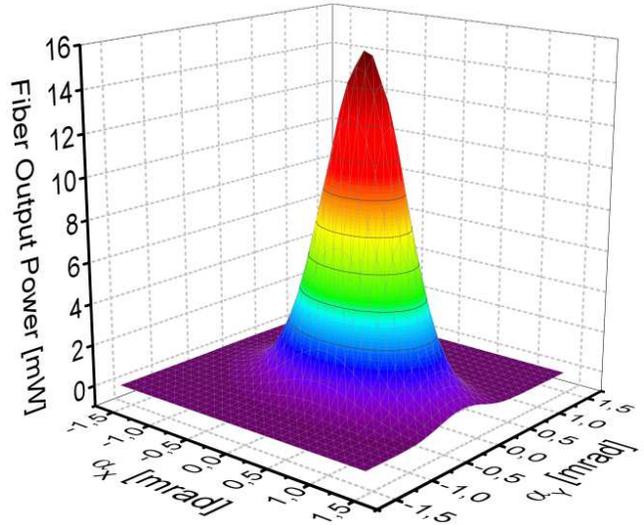}
 \caption{\label{scan_MEF}Typical experimental measurement of the fiber output power when the angle of the piezoelectric mirrors is scanned in the two axes over 2~mrad.}
 \end{figure}

 \begin{figure*}[htb]
 \includegraphics[width=16cm]{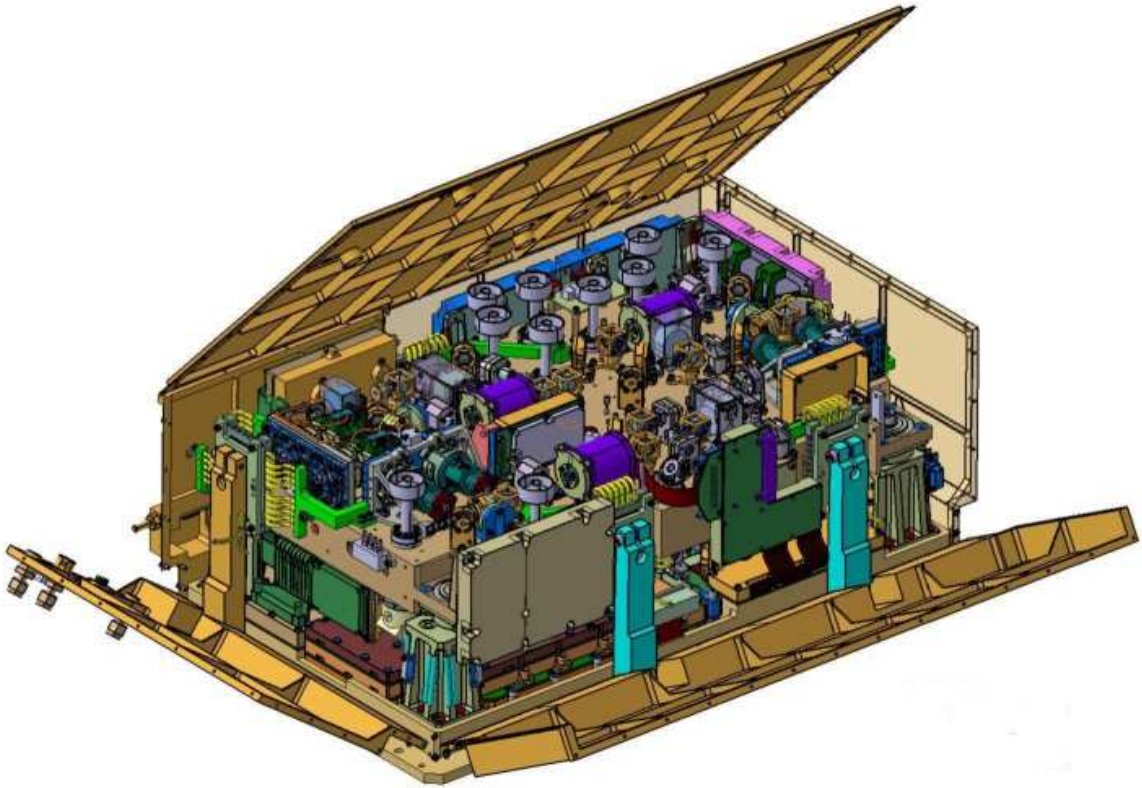}
 \caption{\label{CAO_SL} View of the laser source mechanical design. The dimensions are 532~mm x 335~mm x 198~mm and the mass is 22 kg.}
 \end{figure*}

The tip-tilt mechanisms, depicted on Fig~\ref{MEF}, have been developed by Cedrat Technologies. They are based on 4 piezoelectric elements to generate two independent tilt movements by push-pull actuation. The action of the piezoelectric elements is servo controlled by 4 strain gauges~\cite{leletty2004}. These mechanisms allow to vary the angle of the mirror over a span of $\pm$2~mrad and achieve a long term angular stability below 4~$\mu$rad. The Fig.~\ref{scan_MEF} shows the typical evolution of the fiber output power when the angle is scanned in the two axis.  In non-operational conditions, the tests have proven a compatibility with thermal variations (-50$^{\circ}$C / +75$^{\circ}$C) and random vibrations up to 30~$g_{\mathrm{rms}}$.

\subsection{Laser source assembly and qualification}

The integrated laser source flight model is represented on Fig.~\ref{CAO_SL}. The dimensions are of 532~mm x 335~mm x 198~mm. The optical components are mounted on a double-sided optical bench with dimensions of 450~mm x 284~mm x 40~mm. The bench is fixed on the baseplate through four dampers and its first mechanical resonance frequency is above 150~Hz. The temperature of the bench is regulated at 26 $\pm 0.1^{\circ}$C by means of heaters and Peltier cooler whereas the baseplate temperature can vary from 10 to 33$^{\circ}$C. The laser beams are guided toward the cesium tube by 10 polarization-maintaining fibers (Corning Puremode PM 850 UV/UV-250). The polarization extinction ratio at the output of the fibers is of 30~dB. The power loss induced by the effect of the radiations on the fibers has been estimated to less than 2\% over the mission lifetime~\cite{ott2002}. Ten FC/PC connectors with insertion loss of 1~dB ensure the link with the cesium tube fibers.

The electronics is located on the laser source mounting plate. It includes the power supplies and drivers of all the active elements: laser diode drivers and servo loop, AOM low phase-noise frequency synthesizers, stepper motor drivers and PZT mirror voltage supply. Moreover, an acquisition system allows to digitize the internal parameters of the laser source. An internal FPGA controls the laser source operation and manages the data communication with PHARAO onboard computer through a serial link. It receives 12 logical signals to trigger the different events during clock sequence. The complete system is powered by a 28~V DC line. The total mass of the laser source is of 22~kg and the average electrical consumption is 30~W.

\begin{figure}[!h]
 \includegraphics[width=8.5cm]{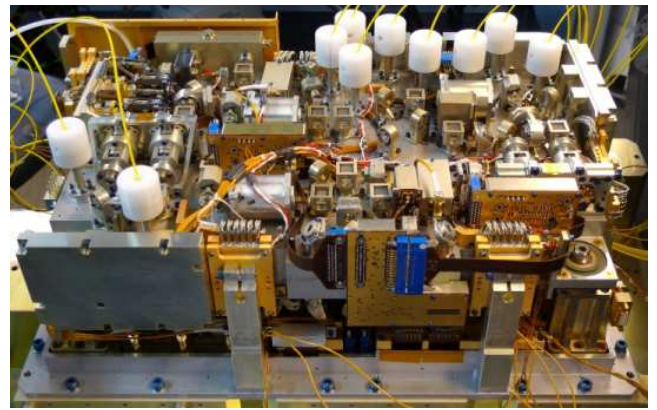}
 \caption{\label{SL_photo}Photograph of the integrated laser source before integration of the cover.}
 \end{figure}

The laser source is one of the most critical parts of a cold atom apparatus. It must keep its performances in terms of delivered optical power, linewidth, frequency stability and spectral agility in a space environment. This system must be reliable enough to undergo high level of vibration and to remain operational over a large range of temperature. The design was validated through the development of a Structural and Thermal Model (STM) and an Engineering Model (EM). The STM was used to perform the qualification tests and the EM was used to validate the functional performances of the laser source.

The complete FM of the laser source, shown on the Fig.~\ref{SL_photo}, was submitted to a set of acceptance tests. Random vibrations were carried out at a level of 6.5~$g_{\mathrm{RMS}}$. Thermal variations in non-operational conditions were made between -32 and +40$^{\circ}$C. Optical performances of the flight model, including output power level and laser spectral purity verifications, were performed after each acceptance tests. Finally, the performances of the flight model were tested in operational conditions in the thermal range of +10$^{\circ}$C / +33.5$^{\circ}$C under vacuum, corresponding to the actual environmental conditions on-board the ISS external support of the European Columbus laboratory. Moreover, Electro-Magnetic Compatibility (EMC) tests were carried out on the EM for qualification and on the FM for acceptance. As the mission will take place in low orbit, the instrument will be submitted to a very low radiation dose of about 700~rad over 3 years. However, the compatibility of each components of the laser system with this dose has been verified.

\section{Laser source performances}
\label{sec_perfo}

\subsection{Frequency noise and RIN measurements}

\begin{figure}[!h]
 \includegraphics[width=8.5cm]{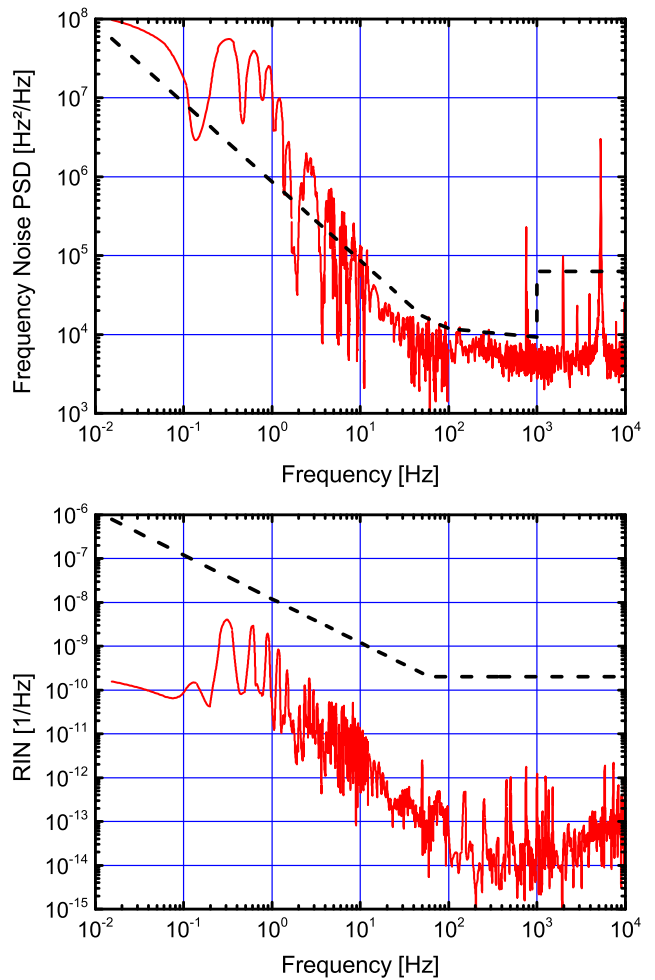}
 \caption{\label{DSP}(Up) Power spectral density of frequency noise of the detection master laser. (Down) Power spectral density of relative intensity noise (RIN) of the detection master laser. The measurement is presented in \textit{red} and the specification is represented in \textit{dashed black}.}
 \end{figure}

The frequency noise and the relative intensity noise (RIN) of the ECDL-M have been measured at the output of the detection fiber. The frequency noise has been measured by a beat note method between the ECDL-M and ECDL-P in closed loop. The power spectral densities (PSD) are given in Fig.~\ref{DSP}. These two features are critical for the performances of the detection system~\cite{leveque2014}. In this part, we present an estimation of the impact of the ECDL-M frequency noise and RIN on the PHARAO clock stability.

\begin{table}[!h]
\caption{\label{tab1} Expected contribution of the relative frequency noise ($\sigma_\nu$) and RIN ($\sigma_{\mathrm{RIN}}$) on the stability of the clock at 1s for 7 given launching velocities $v_0$.}
\begin{tabular}{ccc}
\hline
\hline
 $v_0$(m/s) & $\sigma_\nu$ @1s & $\sigma_{\mathrm{RIN}}$ @1s\\
\hline
0.15 & 6.7 $\times 10^{-15}$ & 1.1 $\times 10^{-16}$\\
0.20 & 1.1 $\times 10^{-14}$ & 1.8 $\times 10^{-16}$\\
0.30 & 1.7 $\times 10^{-14}$ & 2.6 $\times 10^{-16}$\\
0.60 & 2.6 $\times 10^{-14}$ & 3.7 $\times 10^{-16}$\\
1.00 & 3.3 $\times 10^{-14}$ & 4.4 $\times 10^{-16}$\\
3.00 & 5.7 $\times 10^{-14}$ & 6.3 $\times 10^{-16}$\\
3.54 & 6.4 $\times 10^{-14}$ & 6.7 $\times 10^{-16}$\\
\hline
\hline
 \end{tabular}
 \end{table}

The calculation of the influence of the laser frequency noise and RIN on the clock stability is presented in Appendix~\ref{appendix_1}. We carry out the numerical application of this calculation for 7 launching velocities $v_0$. We calculate in each case the contribution of the frequency noise and RIN on the stability of the clock at 1~s. The results are summarized in the Tab.~\ref{tab1}. The frequency and intensity noises of the detection lasers have a very small impact on the clock stability, especially for low launching velocities. For high launching velocities, the clock stability is impacted at a non-limiting level compared to the expected quantum projection noise~\cite{santarelli1999}. Thus, the performances of the instrument are not limited by the frequency and intensity noises of detection laser. Moreover, in the calculation, the contribution of laser frequency noise to the clock stability is largely overestimated ($\Gamma$/2 detuning). In the real case this contribution is then negligible.

\subsection{Cold atom manipulation}

The different PHARAO subsystems have been assembled on the ACES baseplate and connected together. To mimic in-orbit environment, PHARAO is tested in a vacuum chamber ($\sim 2 \times 10^{-6}$~mbar). This significantly changes the thermal behavior of the different sub-systems. Coils surround the chamber to compensate for the earth magnetic field and allow reproducing magnetic orbital variations along the PHARAO main axis.

\begin{figure}[!h]
 \includegraphics[width=8.5cm]{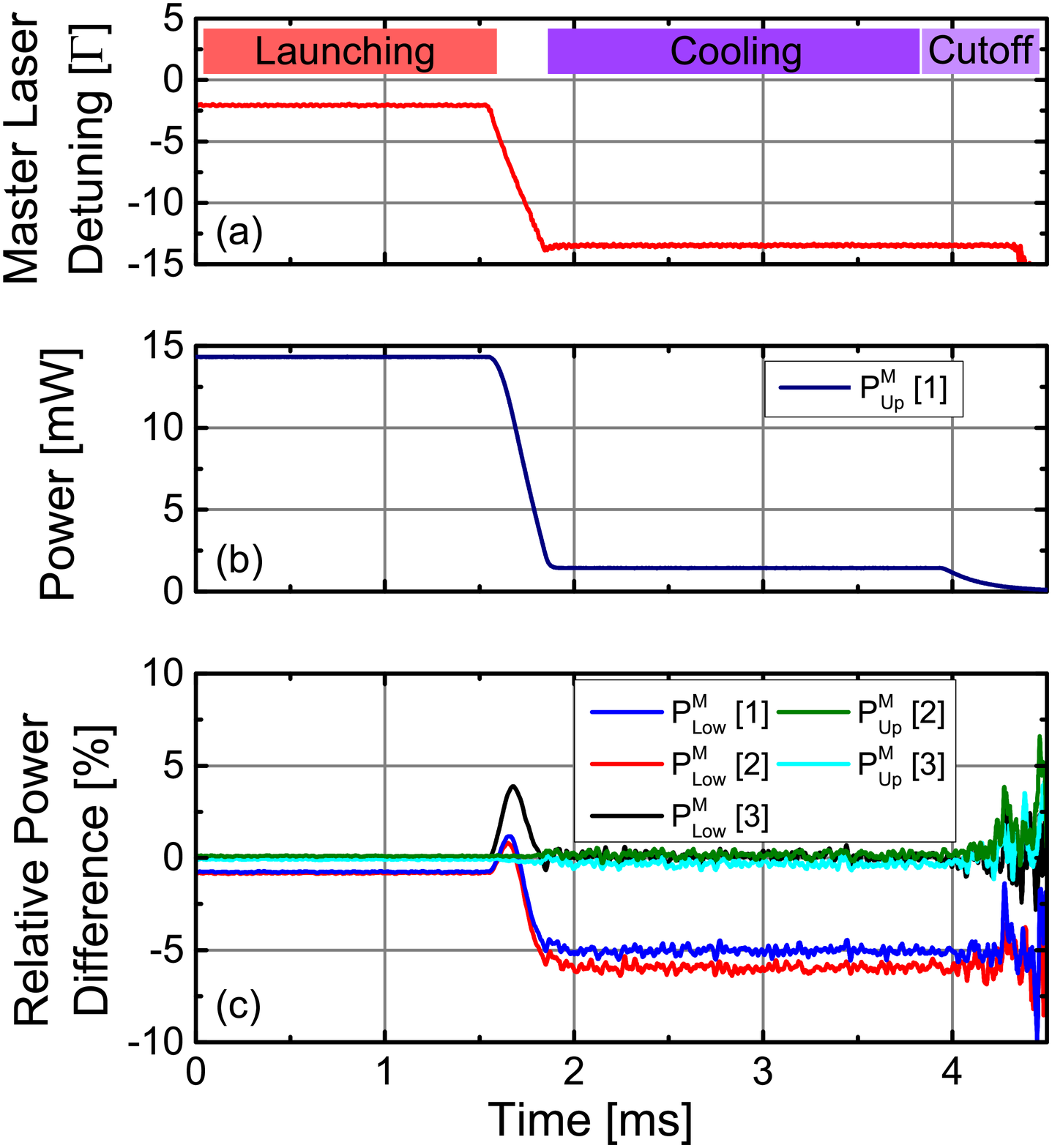}
 \caption{\label{sequence}Measurement of the laser behavior during the preparation sequence. (a) Master laser detuning.  $\Gamma$=5.3 MHz is the natural linewidth of the $^{133}$Cs D2 line. (b) Master power evolution of one upper capture beam ($P^{M}_{Up}[1]$). (c) Relative power difference of the capture beam $P^{M}_{Up}[2,3]$ and $P^{M}_{Low}[1,2,3]$ compared to $P^{M}_{Up}[1]$.}
 \end{figure}

The first step of the testing phase was to tune all the laser source and cesium tube parameters regarding the capture, the cooling, launching, selection, slicing and detection of the atoms. These set of parameters have been optimized for “on-ground” operation in which the atoms are launched upwards at 3.54~m/s. In this specific configuration, the velocity of the atomic cloud decreases along the cesium tube due to the gravity. The atom velocity at the level of the selection and detection system is then respectively of 3.2 and 1.3~m/s. Some of the parameters determined during these tests will need readjustment once in orbit (e.g. launching velocity can be changed at will from 0.05 to 5~m/s) owing to the peculiar behavior of the cold atom cloud in micro-gravity.

First, the parameters of each phase of the preparation sequence, represented on the Fig.~\ref{preparation}, have been set up in order to get the largest number of atoms at the the lowest temperature. The Fig.~\ref{sequence} presents the master laser detuning and the capture beam powers during a typical preparation sequence. The laser detuning has been measured with a frequency to voltage converter by a beat note technique. The measurement shows that the capture beam powers remain balanced within 5\% during the preparation sequence. The temperature and the number of atoms have been optimized using the time-of-flight signal recorded by the detection system. As the number of detected atoms is large and can saturate the detected signal, the loading time is set to 200~ms and the capture laser power is decreased to 5~mW/beam for the on-ground test configuration. The number of atoms is determined from the fluorescence value. The current set of parameters enables to capture up to $5 \times 10^7$ atoms at a temperature of about 1~$\mu$K.

\begin{figure}[!h]
 \includegraphics[width=8.5cm]{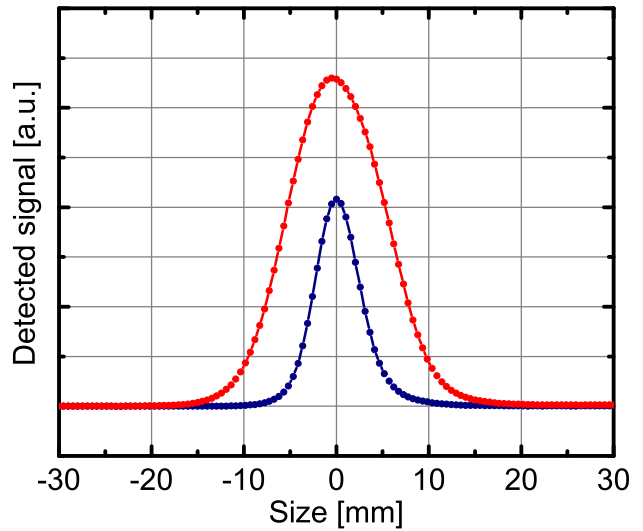}
 \caption{\label{TOF}Detected signal time-of-flight of the atoms. The atomic distribution at the detection is represented without (red points) and with slicing (blue points).}
 \end{figure}

Second, the power of the pumping and pushing selection beam have been optimized. The power of the pushing beam $P^{M}_{Sel}$ has been set to 2.8~mW in order to prepare the cloud in a pure F=3 quantum state. Then, the efficiency of the pumping beam has been verified by measuring the remaining number of detected atoms in $F$=3 after the selection.  For 0.3~mW of pump power, we reach the minimum of atoms (1\%) corresponding to the residual atoms in $F$=3 after the cooling phase. This parameters allow an efficient slicing of the atomic cloud, depicted in the  Fig.~\ref{TOF}.

\begin{figure*}[htb]
 \includegraphics[width=\linewidth]{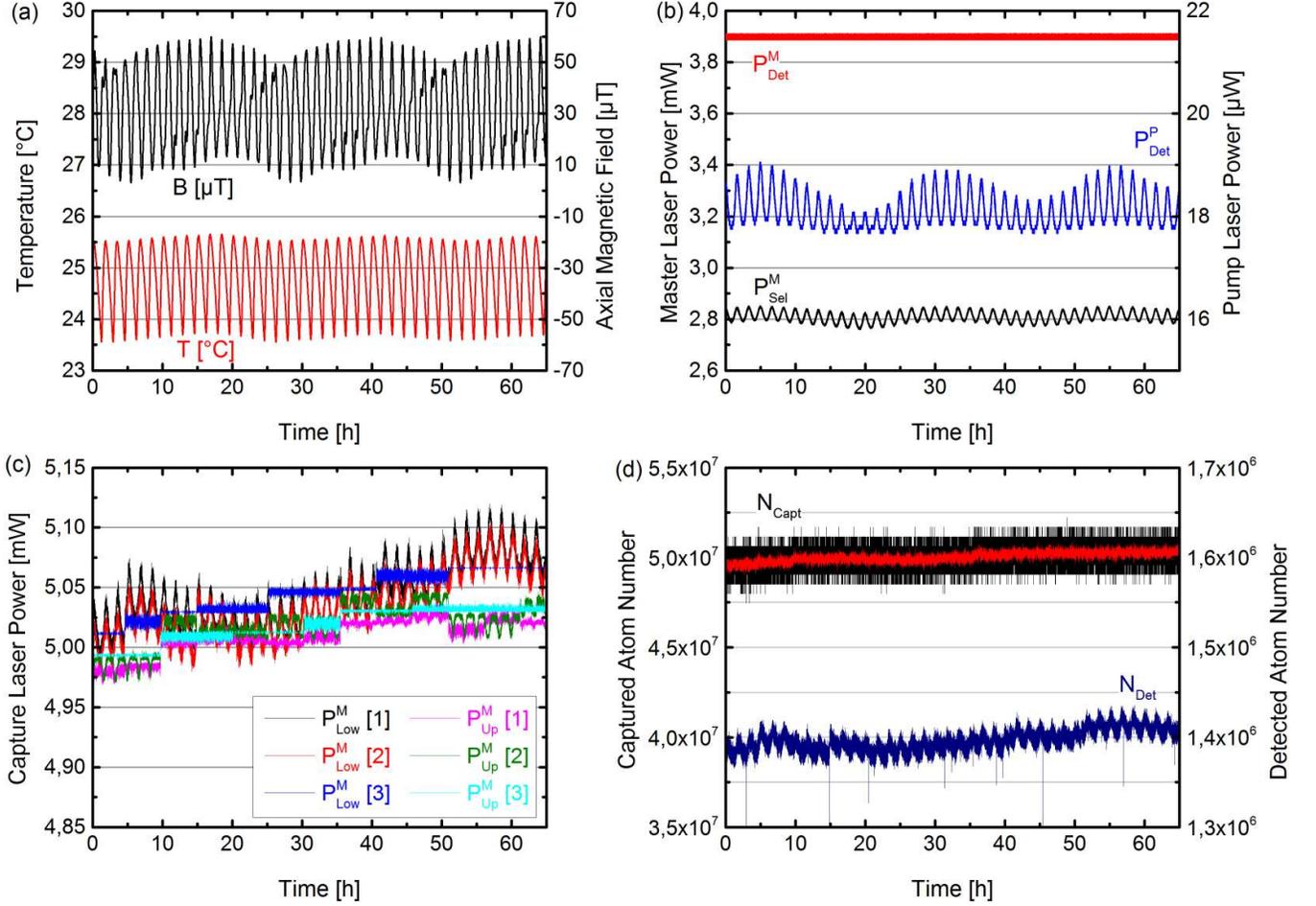}
 \caption{\label{stab_sl} Fiber output power stability during the orbital cycling tests. (a) \textit{red}: Temperature at the level of the laser source baseplate. \textit{black}: Magnetic field in the launching direction measured inside the cesium tube. (b) Stability of the master detection power ($P^{M}_{Det}$, \textit{red}), of the selection master power ($P^{M}_{Sel}$, \textit{black}) and of the pump detection power ($P^{P}_{Det}$, \textit{blue}). (c) Stability of the upper ($P^{M}_{Up}[1,2,3]$) and lower ($P^{M}_{Low}[1,2,3]$) master power during the capture phase. The data are averaged over 100~s. (d) Stability of the captured ($N_{Capt}$) and detected ($N_{Det}$, \textit{blue}) atom number. The captured atom number data are presented shot-to-shot (\textit{black}) and averaged over 100~s (\textit{red}).}
 \end{figure*}

Finally, the detection system has been tested. The master power $P^{M}_{Det}$ has been set to 3.9~mW in order to guarantee the efficiency of the pushing beam. A pumping power $P^{P}_{Det}$ of 20~$\mu$W ensures an optimal operation of the detection system in the test configuration where the atoms velocity is 1.3~m/s at the detection level. The detection pump power can be adjusted up to 94~$\mathrm{\mu}$W. The power margin is then sufficient for higher launching velocities. In standard operation, due to magnetic level selection and clipping by the microwave cavity apertures, we detect typically $1.4 \times 10^6$ atoms. The size of the detected cloud is of 16~mm.

\subsection{Output power stability}

In order to test the reliability of the PHARAO clock in operational conditions, the FM was submitted to orbital cycling tests. During this period of 65 hours, the payload is submitted to thermal and magnetic oscillations representative of the environmental conditions on-board the ISS. The laser source behavior during the orbital cycling tests is presented on the Fig.~\ref{stab_sl}. The Fig.~\ref{stab_sl}(a) depicts the evolution of the temperature at the level of the laser source baseplate during the test. The optical powers are measured inside the cesium tube (\textit{i.e.} at the output of the optical fibers).

The capture master laser power is given on the Fig.~\ref{stab_sl}(c) for the upper ($P^{M}_{Up}[1,2,3]$) and lower ($P^{M}_{Low}[1,2,3]$) beam triplets. First, we see that the relative power stability of the 6 beams remains within 1\%. This is due to the small power instability correlated to the temperature.  The measurements exhibit a long term drift of the laser power (0.3\%/day). This small drift was caused by an anomaly in the recalibration software of the piezoelectric mirrors which has been corrected after the test. The result of the test is compliant with the power stability specification of $\pm$3\%. The four capture pump laser powers, which are not critical for the number of captured atoms, have also been measured and remain constant over the entire test, without apparent drifts. The number of captured and detected atoms, given on Fig.~\ref{stab_sl}(d), are perfectly correlated to the drift of the capture laser power. These variations will be canceled by the correction of the software anomaly.

The laser powers in the selection and detection systems are represented on the Fig.~\ref{stab_sl}(b). The power in the selection master laser ($P^{M}_{Sel}$), which is not optically servo controlled, is stable at the level of 3\%. A correlation analysis shows that this power has a thermal sensibility of 1\%/$^{\circ}$C. This behavior is then compliant with the power stability specification of $\pm$3\%. The power in the detection master laser ($P^{M}_{Det}$), which is optically servo controlled, shows a stability below the measurement sensitivity ($<$0.5\%). The power in the detection pump laser ($P^{P}_{Det}$), which is free running, shows a 15\% stability over the 65 hours considered. The measurement exhibits a small sensitivity to the temperature variations at the level of 3\%/$^{\circ}$C. It is slightly larger than the initial specification ($\pm$5\%) but the stability of this power is less critical than expected for the detection process since no significant influence on the detected atom number has been observed. Nevertheless, a periodic adjustment of $P^{P}_{Det}$ has been added after the tests in order to prevent any long-term drift.

\section{Conclusion}

In summary, the first tests of the laser source flight model have confirmed a nominal behavior, even in orbital representative environment. We have presented the design of the laser source flight model of the PHARAO cold atom clock. The main functions of this system and the key technologies developed for this mission  have been described. The parameters of the laser source for cold atom manipulation have been determined. The characteristics of the laser source, and their impact on the clock performances, have been verified in operational conditions. In particular, the noise of the laser source has been characterized and its influence on the stability of the clock has been shown to be negligible. The environmental acceptance tests have been carried out successfully. The stability of the output powers in orbital condition has been measured and is compatible with a nominal operation of the PHARAO clock. The complete PHARAO flight model has been assembled and extensively tested at CNES. It has demonstrated a level of stability of $3.2 \times 10^{-13} \tau^{-1/2}$ expected for ground operation. Finally, ACES will be assembled in 2015 for a launch envisioned in 2017 by SpaceX. 

To a great extent, the flight model presented in this work is the first laser system designed for spaceborne cold atom manipulation. The space qualified technologies developed in the frame of this project are easily transposable to other wavelength and pave the way for future cold atom space missions including optical clocks~\cite{schiller2012} and atom interferometers~\cite{schuldt2015,carraz2014}.

\begin{acknowledgments}
The laser source has been manufactured and qualified for space by Sodern under CNES contract. We would like to thank C. Escande, P. Larivi\`ere and S. Beraud for their contribution to the integration and test of the laser source inside PHARAO. We thank all the members of the PHARAO industrial team at CNES: Ph. Chatard, C. M. de Graeve, S. Tellier, C. Stepien, L. Fonta, T. Basquin, A. Ratsimandresy, S. Julien and E. Leynia de la Jarrige for their outstanding work in assembling and testing the PHARAO flight model.
\end{acknowledgments}


\appendix

\section {Influence of the laser noise on the clock stability}
\label{appendix_1}

In this part, we present an estimation of the impact of the ECDL-M frequency noise and RIN on the PHARAO clock stability. The transition probability $P$ of the atoms at the output of the clock is determined by measuring the ratio between the number of atoms in each ground state by a fluorescence method. The atomic cloud passes through two thin detection beams. The fluorescence signals emitted by spontaneous emission are collected on two photodiodes. The areas of these two time-of-flight signals, $A_{|e\rangle}$ and $A_{|f\rangle}$, are proportional to the number of atoms in each ground state. The relative frequency and intensity fluctuations of the ECDL-M induce variations on the spontaneous emission rate during the detection process. This effect results in a perturbation on the measurement of the atom number present in the detection beams. We denote by $f (t)$ the ratio between the number of atoms present in the first detection beam at a time t and the corresponding total time-of-flight area $A_{|e\rangle}$. At the time $t + \Delta t$, the atoms pass through the second detection beam, leading to fluctuations in the area of flight time $A_{|f\rangle}$. We consider a Gaussian type profile for the time-of-flight signal:

\begin{equation}
f(t) = \frac{1}{\sqrt{2 \pi} w} \exp \left( - \frac{t^2}{2 w^2} \right)
\end{equation}

The temporal width of the time-of-flight $w$ is connected to the initial width of the atomic cloud ($w_x = 5.52 \times 10^{-3}$~m), the launching velocity ($v_0$) and the width of the velocity distribution of the atomic cloud ($w_v = 8 \times 10^{-3}$~m/s). The time-of-flight signal is the result of the convolution between the initial width of the cloud and the width of the velocity distribution weighted by the total flight time $t_{\mathrm{tof}}$ of the atoms between the capture and the detection. The temporal width of the flight time is therefore given by:

\begin{equation}
w = \frac{1}{v_0} \sqrt{w_x^2+\left({w_v t_{\mathrm{tof}}}\right)^2}
\end{equation}

We therefore neglect the width of the probe compared to the size of the atom cloud at the time of the detection. This assumption is particularly valid when the launching velocity is low. The total flight time $t_{\mathrm{tof}}$ is connected to the launching velocity $v_0$ and the length between the capture zone and the detection zone ($L$=0.54~m). We therefore have $t_{\mathrm{tof}} = L/v_0$.

The fluctuation of the time of flight area $\delta A_{|f\rangle,|e\rangle}$ for each ground state $|f\rangle$ and $|e\rangle$ is given by:

\begin{equation}
\delta A_{|f\rangle,|e\rangle} = A_{|f\rangle,|e\rangle} \times \int_{-\infty}^{+\infty}{\delta \gamma (t) f(t) dt}
\end{equation}

Where $\delta \gamma (t)$ is the fluctuation of the spontaneous emission rate. The transition probability $P$ is measured by:

\begin{equation}
P=\frac{A_{|e\rangle}}{A_{|e\rangle}+A_{|f\rangle}}
\end{equation}

This probability is then affected by a fluctuation $\delta P$ which can be expressed, in the particular case of mid-fringe measurement, as:

\begin{equation}
\delta P=\frac{1}{4}\int_{-\infty}^{+\infty}{\delta \gamma (t) \left[{ f(t) - f(t- \Delta t) }\right] dt }
\end{equation}

The standard deviation of these fluctuations is then given by:

\begin{equation}
\sigma_{\delta P}^2 = \int_{0}^{+\infty}{S_{\delta \gamma} (f) \left|{H_{\mathrm{det}}(f) }\right|^2 df }
\end{equation}

Where $S_{\delta \gamma} (f)$ is the PSD of the fluctuations of the spontaneous emission and $H_{\mathrm{det}}(f)$ is the transfer function of the detection system:

\begin{equation}
H_{\mathrm{det}}(f)=\frac{1}{4} \left({1 - e^{-2 i \pi f \Delta t}}\right) \times e^{-2 \pi^2 f^2 w^2}
\end{equation}

The PSD of the spontaneous emission rate fluctuations $S_{\delta \gamma} (f)$ can be related to the frequency fluctuations $S_{\nu} (f)$ and to the relative intensity noise $S_{\mathrm{RIN}} (f)$ of the ECDL. In this calculation, we consider the case of unsaturated atoms and of a $\Gamma$/2 red detuned detection laser, thereby maximizing the sensitivity to the laser frequency fluctuations. It is given, at the first order by:

\begin{equation}
S_{\delta \gamma} (f) = \left({ \frac{4 \pi \nu_{\mathrm{laser}}}{\Gamma} }\right)^2 \times S_{\nu} (f) + S_{\mathrm{RIN}} (f)
\end{equation}

The final contribution of the laser noise to the stability of the clock ($\sigma_y$) is given by:

\begin{equation}
\sigma_y (\tau) = \frac{2}{\pi Q_{\mathrm{at}}} \times \sigma_{\delta P} \times \sqrt{\frac{T_c}{\tau}  }
\end{equation}

Where $\tau$ is the integration time, $T_c$ is the cycling time of the clock and $ Q_{\mathrm{at}}$ is the atomic quality factor.

\end{document}